# Toward a Gigabit Wireless Communications System

L. Rakotondrainibe, Y. Kokar, G. Zaharia and G. El Zein

Institute of Electronics and Telecommunications of Rennes, IETR - UMR CNRS 6164,
INSA, 20 Avenue des Buttes de Coesmes, CS 14315, 35043 Rennes cedex, France
*Lahatra.Rakotondrainibe@insa-rennes.fr*

*Abstract*: This paper presents the design and the realization of a hybrid wireless Gigabit Ethernet indoor communications system operating at 60 GHz. As the 60 GHz radio link operates only in a single-room configuration, an additional Radio over Fiber (RoF) link is used to ensure the communications within all the rooms of a residential environment. The system uses low complexity baseband processing modules. A byte synchronization technique is designed to provide a high value of the preamble detection probability and a very small value of the false detection probability. Conventional RS (255, 239) encoder and decoder are used for channel forward error correction (FEC). The FEC parameters are determined by the trade-off between higher coding gain and hardware complexity. The results of bit error rate measurements at 875 Mbps are presented for various antennas configurations.

## 1. Introduction

The massive use of high quality multimedia applications such as high definition video streaming, file transfer and wireless Gigabit Ethernet explains the need of 1 Gbps overall throughput. One of the most promising solutions to achieve a gigabit class wireless link is to use millimeter-waves (MMW) for the carrier frequency. Due to the large propagation and penetration losses, 60 GHz Wireless Personal Area Networks (WPANs) are primarily intended for use in short range and single room environments. In addition to the high data rates that can be achieved in this frequency band, the radio waves propagation at 60 GHz has specific characteristics offering many other benefits such as high security and frequency re-use.

High frequency and even MMW analog communication circuits, which were traditionally built on more expensive technologies such as bipolar or Gallium Arsenide (GaAs), are progressively implemented on CMOS. Table 1 summarizes some experimental previous works about the realized transceiver operating at 60 GHz [1], [2]. Different architectures have been analyzed in order to develop new MMW communication systems for commercial applications [3]-[7]. The selection of a modulation scheme is a primary consideration for any wireless system design and has a large impact on the system complexity and power consumption as well as issues such as amplifier linearity, and oscillator phase noise. Moreover, to determine the appropriate system design at 60 GHz, wireless channel characteristics must be well understood. This includes path loss, material attenuation, multipath effects, and antennas [3], [4]. Hence, other considerations such as synchronization, coding/error correction and equalization must be taking into consideration for the overall system design.

*Table 1.* Summary of experimental 60 GHz radio systems

|  | **Wigwam (2005)** | **IBM (2006)** | **NEC (2002)** | **Motorola (2004)** |
|---|---|---|---|---|
| Technologies | 0.25 µm SiGe BiCMOS | 0.15 µm AlGaAs/ InGaAs HJFET | 0.13 µm SiGe BiCMOS | AsGa |
| Modulation | OFDM /QPSK | OFDM/QPSK ASK, MSK | ASK | OOK |
| Keywords | LNA, PA, Mixer, PLL | LNA, PA, Mixer, LO | Mod/demod Filter, LNA, doubler | LNA, PA, Mixer, Multiplier |
| Intermediate Frequency | 4,5-5,5 GHz | 9 GHz | - | - |
| Performance | 250 Mbps 1 m | 630 Mbps, OFDM-10m 1 Gbps, ASK-1m | 1,25 Gbps 7 m | 3,5 Gbps 3 m |

This paper proposes a hybrid communication system derived from the simplified IEEE 802.15.3c physical (PHY) layer [8] to ensure near 1 Gbps data rate on the air interface. The first system application in a point-to-point configuration is the high-speed file transfer. The system must operate in indoor, line-of-sight (LOS) domestic environments.

The rest of this paper is organized as follows. Sections 2 and 3 describe respectively the transmitter and the receiver block diagrams. In these sections, the radiofrequency (RF) architecture is first presented. Then, the baseband blocks are described. The byte/frame synchronization method also is discussed. Measurement results and analysis are respectively presented in sections 4 and 5. Section 6 concludes the work.

## 2. Transmitter design

Figure 1 shows the block diagram of the realized transmitter (Tx) system.

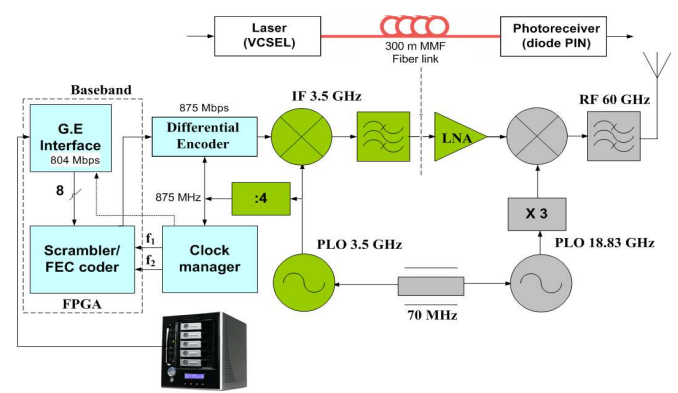

**Figure 1.** 60 GHz Wireless Gigabit Ethernet transmitter

The system uses a single carrier architecture based on a Differential Encoded Binary Phase Shift Keying (DBPSK) modulation. A differential encoder is used before the BPSK modulation to remove the phase ambiguity knowing that a differential demodulation is used at the receiver (Rx). Compared to higher order constellations or OFDM systems, this system is more resistant to phase noise and power amplifier (PA) non-linearities. OFDM requires large back-off for PA, high stability and low phase noise for local oscillators. Furthermore, the implementation of a single-carrier system is simple.

### 2.1 Radiofrequency architecture

After channel coding and scrambling, the input data are differentially encoded using logic circuits (PECL). The differential encoder performs the delayed modulo-2 addition of the input data bit $x_n$ with the output bit $y_n$ as in (1).

$$y_{n+1} = x_n \oplus y_n \quad (1)$$

The obtained data are used to modulate an intermediate frequency (IF) carrier generated by a 3.5 GHz phase locked oscillator (PLO) with a 70 MHz external reference. The IF modulated signal is fed into a band-pass filter (BPF) with 2 GHz bandwidth and transmitted through a 300 m optical fiber link. This IF signal is used to modulate the current of the Vertical Cavity Surface Emitting Laser (VCSEL) operating at 850 nm. In order to avoid signal distortions, the power of the VCSEL input signal must not exceed -3 dBm. A 300 m multimode fiber with a bandwidth-length product of about 4000 MHz.km is enough to cover the bandwidth and the distance range. After transmission via the fiber, the optical signal is converted to an electrical signal by a PIN diode and amplified. The overall RoF link has 0 dB gain and an 8 GHz bandwidth.

Following the RoF link, the IF signal is sent to the RF block. This block is composed of a mixer, a frequency tripler, a PLO at 18.83 GHz and a band-pass filter (59-61 GHz). The frequency of the local oscillator is obtained with an 18.83 GHz PLO with a 70 MHz reference and a frequency tripler. The phase noise of the 18.83 GHz PLO signal is about –110 dBc/Hz at 10 kHz off-carrier. The upper sideband is selected using a BPF. The BPF prevents the spill-over into adjacent channels and removes out-of-band spurious signals caused by the modulator. The 0 dBm obtained signal is fed into the 22.4 dBi horn antenna with 10° vertical half-power beamwidth and 12° horizontal half-power beamwidth (HPBW).

### 2.2 Baseband architecture

The Gigabit-Ethernet interface is used to connect a home server to the 60 GHz wireless link with around 800 Mbps bit rate, as shown in Figure 2.

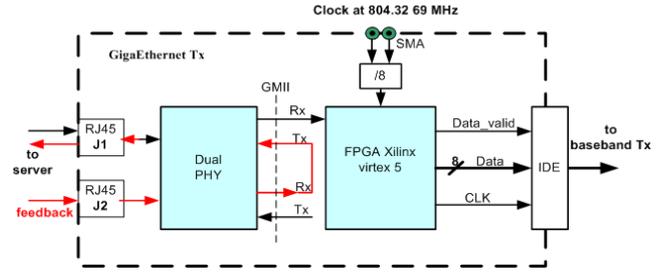

**Figure 2.** Transmitter Gigabit Ethernet interface

The Gigabit Media Independent Interface is an interface between the Media Access Control (MAC) device and the physical layer (PHY). This interface defines data rates up to 1 Gbps, implemented using byte data interface at 125 MHz. However, this frequency is different from the clock data (104.54 MHz) generated by a clock manager, as explained latter. This clock asynchronism provides data packet loss. In order to avoid jitter, a programmable circuit (FPGA) is used as part of buffers memory.

The transmitter baseband architecture is shown in Figure 3.

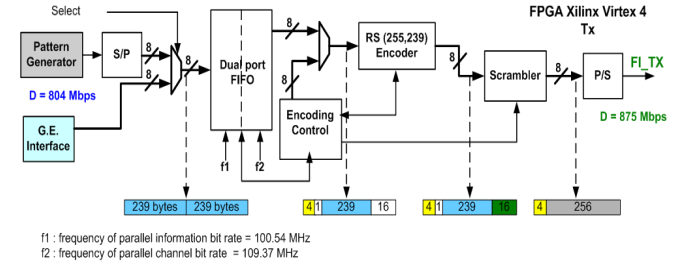

**Figure 3.** Transmitter baseband architecture

The data bytes from the Gigabit Ethernet interface are directly transfered into the dual port FIFO memory (due to the byte streams of the RS encoder). Otherwise, the input bit stream is paralellized by a serial-to-parallel (S/P) converter.

The frame format consists of 4 preamble bytes, 239 data burst bytes, 16 check bytes and a "dummy byte" as shown in Figure 4. The dummy byte is added besides 4 preamble bytes to obtain a multiple of 4 coded data length useful for the scrambling operation.

A known preamble is sent at the beginning of each frame in order to achieve the frame synchronization at the receiver. The used preamble is a Pseudo-Noise (PN) sequence of 31 bits + 1 bit to provide 4 preamble bytes.

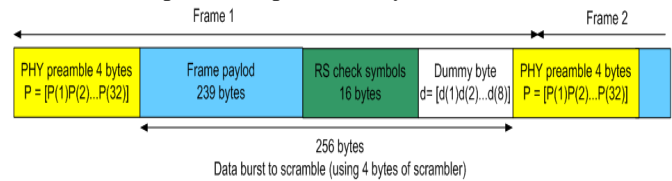

**Figure 4.** Frame structure

Hence, owing to the frame structure and the byte operation, two different clock frequencies $f_1$ and $f_2$ are used in the baseband blocks:

$$f_1 = \frac{F_1}{8} = 100.54 \text{ MHz}, \quad f_2 = \frac{F_2}{8} = 109.37 \text{ MHz}.$$

where:

$$F_2 = \frac{IF}{4} = 875 \text{ MHz} \quad \text{and} \quad \frac{F_1}{F_2} = \frac{239}{260}.$$

The frame structure is obtained as follows: byte streams are written into the dual port FIFO memory at 100.54 MHz. The FIFO memory has been set up to use two different clock frequencies for writing at $f_1$ and reading at $f_2$. Therefore, reading can be started when the FIFO memory is half-full in order to avoid the under/overflow of the FIFO memory. The encoding control generates 4 preamble bytes (not processed by the RS encoder) and reads 239 bytes. The RS encoder reads one byte every clock period. After 239 clock periods, the encoding control interrupts the bytes transfer during 17 clock periods. During this interruption, 4 preamble bytes are added; the encoder takes 239 data bytes and appends 16 control bytes to make a code word of 255 bytes. In addition, the encoding control adds the dummy byte for the balanced scrambling operation. In all, the number of data coded bytes (excepting the 4 preamble bytes) is 256, a multiple of 4.

The additional dummy byte $d = [d(1)\ d(2)\ \ldots\ d(8)]$ is determined in order to obtain the lowest value of the maximum cross-correlation between the preamble P and T, given by (2) and (3), where:

$$P = [P(1)\ P(2)\ \ldots\ P(32)] \quad (2)$$

and

$$T = [d(9-i)\ldots d(8)\ d(1)\ P(1)\ P(2)\ \ldots\ P(32-i)],\ 1 \le i \le 8.\quad (3)$$

We note $k = d(8)*2^7+d(7)*2^6+\ldots+d(2)*2+d(1)$, $0 \le k \le 255$. Therefore, for each value of k, a different frame header is obtained. The maximum value of the cross-correlation between P and T is computed:

$$\text{Mcor}(k) = \max_i \left\{ \text{sum}(\overline{P \oplus T}) \right\}. \quad (4)$$

This value is presented in Figure 5 as a function of k. The minimum of the maximum cross-correlation value is obtained for several values of k.

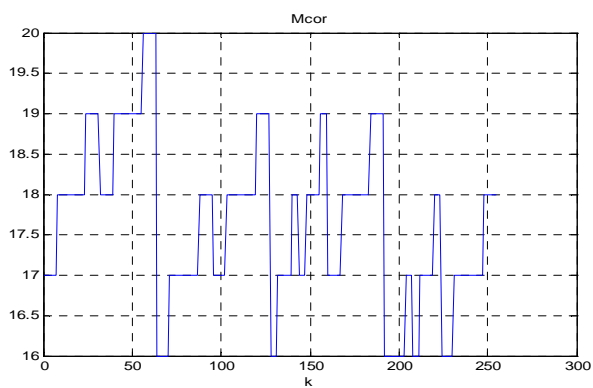

**Figure 5.** Maximum correlation between P and T

For each value of k, the cross-correlation is represented for $1 \le i \le 8$. All the results are analyzed to identify the best dummy byte which gives the lowest cross-correlation values. The best result is obtained for k = 64 which gives d = [0 0 0 0 0 0 1 0]. For k = 64, the cross-correlation for different possible shifts ($1 \le i \le 8$) is shown in Figure 6.

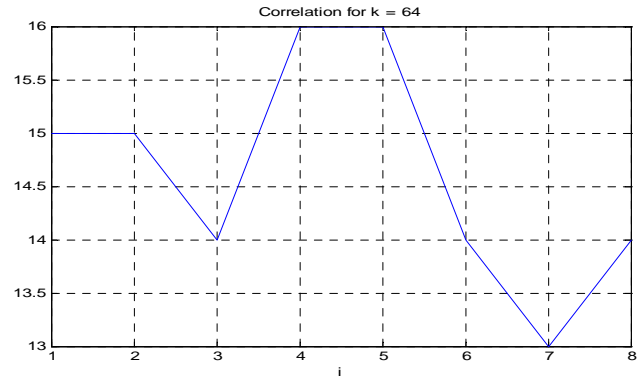

**Figure 6.** Cross-correlation between P and T for k = 64

The maximum cross-correlation is equal to 16. This value is obtained twice; all the other values are smaller. Therefore, this optimal byte gives the smallest value of the PHY preamble false detection probability.

Long sequences of '0' or '1' must be avoided for reliable operation of Clock and Data Recovery (CDR) circuit at the receiver. Therefore, the use of a scrambler at the transmitter and a descrambler at the receiver is essential. The scrambler is a PN-sequence of 31 bits + 1 bit to obtain 4 bytes. This scrambler is chosen in order to provide the lowest cross-correlation values between the received data and the 4 bytes PHY preamble. This method reduces the number of false detections of the PHY preamble within the scrambled data. The received byte stream is finally parallel-to-serial (P/S) converted just before the differential encoder.

## 3. Receiver design

Figure 7 shows the block diagram of the realized receiver system. The system architecture uses a simple non coherent demodulation at the receiver.

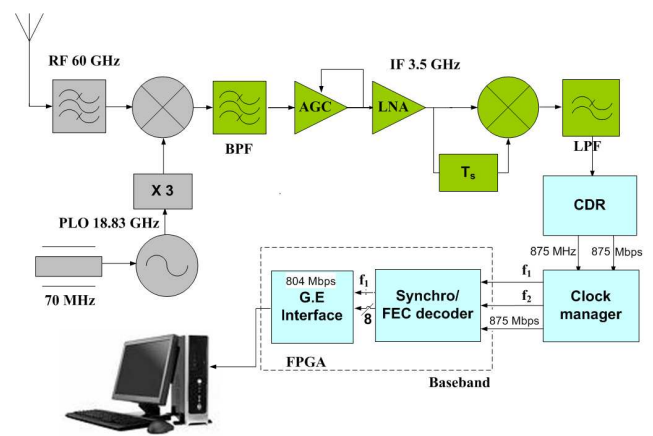

**Figure 7.** 60 GHz Wireless Gigabit Ethernet receiver

### 3.1 Radio frequency architecture

The input band-pass filter, after the Rx antenna, removes the out-of-band noise and adjacent channels interference. The RF signal obtained at the filter output is down-converted to IF = 3.5 GHz and fed into a band-pass filter with 2 GHz bandwidth. An Automatic Gain Control (AGC) with a dynamic range of 20 dB is used to ensure a quasi-constant power level at the demodulator input. A Low Noise Amplifier (LNA) with a gain of 40 dB is used to achieve sufficient gain. The differential demodulation enables the coded signal to be demodulated and decoded.

Compared to a coherent demodulation, this method is less efficient in additive white Gaussian noise (AWGN) channel. However, in the presence of frequency selective propagation channel, the differential demodulation is more robust to multi-path propagation which induces intersymbol interference (ISI). Indeed, the differential demodulator is realized with a delay line (Ts = 1.14 ns) and a mixer. Then, due to the product of two consecutive symbols, the rate between the main lobe and the second lobes of the impulse response of the channel is increased. This operation improves the Signal to Noise + Interference ratio (SNIR). Following the differential demodulator, a Low-Pass Filter (LPF) with 1 GHz cut-off frequency removes the high-frequency components of the obtained signal.

### 3.2 Baseband architecture

Figure 8 shows the receiver baseband architecture. The serial data (after the CDR) are converted into byte streams.

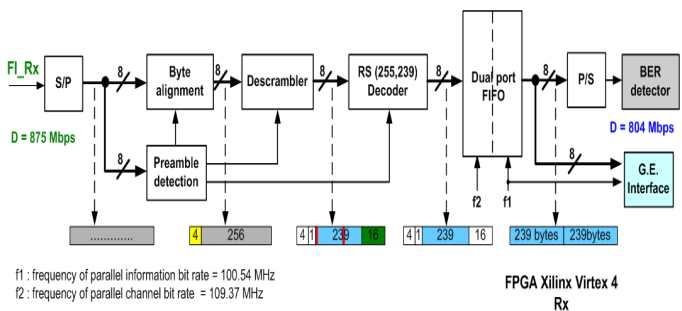

**Figure 8.** Receiver baseband architecture

Figure 9 shows the physical preamble detection architecture.

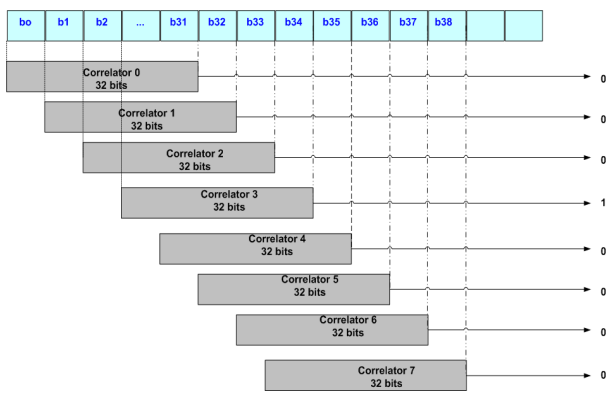

**Figure 9.** PHY Preamble detection architecture

The preamble detection is based on the cross-correlation between 32 successive received bits and the internal 32 bits PHY preamble. In fact, this preamble detection is performed within a FPGA by using a byte stream in order to reduce the data clock frequency. Because the serial-to-parallel conversion realized by the multigigabit transceiver of the FPGA is arbitrary, a byte-synchronization must be performed. Therefore, 8 identical correlators are used. In addition, each correlator must analyse a 1-bit shifted sequence of 32 bits. Hence, the preamble detection is performed with 32 + 7 = 39 bits (+ 7 because of different possible shifts of a byte). In all, there are 8 corrrelators of 32 bits within each correlator-bank, as shown in Figure 10.

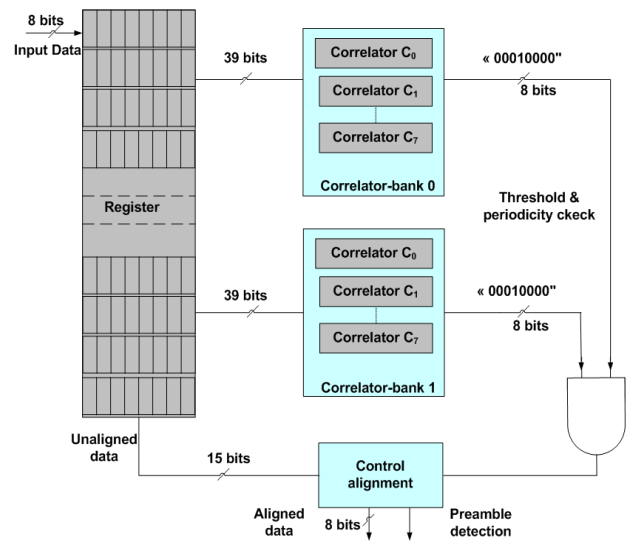

**Figure 10.** Preamble detection with periodicity and control alignment

Further, in order to obtain a very low preamble miss detection probability, the periodical repetition of the PHY preamble is taken into consideration. The decision is made from two successive PHY preambles (preamble1 + 256 data bytes + preamble2) stored in a register. If the preamble detection is indicated in each correlator-bank by the same correlator $C_k$, the operation is validated.

Each value of the correlation between the preamble and the received data (32 bits) is compared to the threshold to be determined. Threshold setting plays an important role because the performance of the proposed system is based on the reliable byte/frame synchronization. The frame synchronization performance is characterized by the miss detection probability $P_m$, the false alarm probability $P_f$ and the channel error probability p, as described in [9]. Figure 11 shows the miss detection probability of the PHY preamble as a function of p. The maximum setting threshold S = 32 is equal to the maximum cross-correlation between the PHY preamble (32 bits) and the received data (32 bits). This maximum cross-correlation is obtained when the preamble of the received frame has no error.

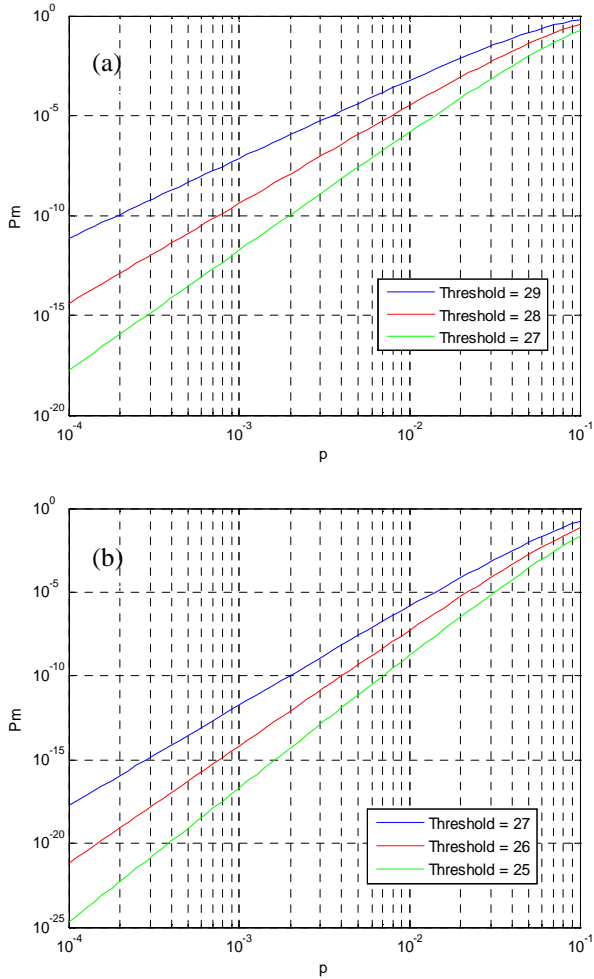

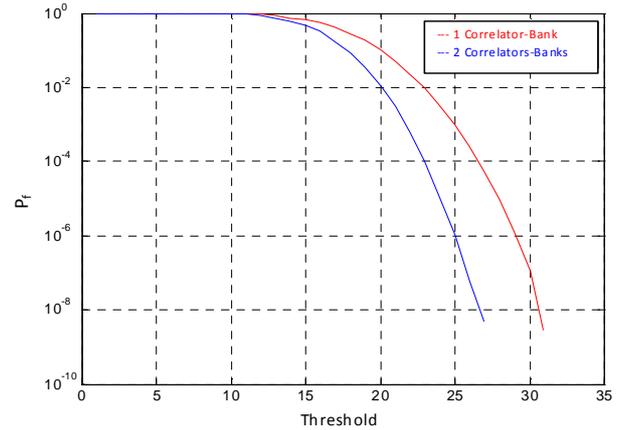

**Figure 12.** False alarm probability of PHY preamble detection with one/two blocks of correlators

After the byte alignment and the preamble detection, the descrambler performs the modulo-2 addition between 4 successive received data bytes and the 4 bytes scrambler. The RS decoder processes the descrambled bytes and attempts to correct the errors. The RS (255, 239) decoder can correct up to 8 erroneous bytes and operates at a high clock frequency ($f_2$ = 109.37 MHz). The byte streams obtained at the decoder output are written discontinuously in the FIFO dual port memory with the clock frequency $f_2$. The other frequency $f_1$ = 100.54 MHz reads out continuously the data bytes stored in the FIFO dual port memory. Then, the byte streams are finally transmitted to the Gigabit Ethernet interface, as depicted in Figure 13.

**Figure 11.** PHY preamble miss-detection probability

The byte/frame synchronization also depends on the false alarm events. Occasionally, due to the random values of the binary transmitted data, the architecture shown in Figure 10 can falsely declare that the preamble is detected. For this reason, the threshold is chosen in order to obtain the best trade-off between a high value of the detection probability and a very small value of the false alarm probability. A false alarm is declared when the same correlators $C_k$ of the two correlator-banks indicate the detection of the preamble within two successive bursts of 256 data bytes. Figure 12 shows the preamble false alarm probability for $p = 10^{-2}$.

Thus, by taking into account the periodical repetition of the preamble, we obtain:

- for $p = 10^{-2}$ and S = 27, Pm ≈ $10^{-6}$ and Pf ≈ $2*10^{-10}$;
- for $p = 10^{-2}$ and S = 28, Pm ≈ $3*10^{-5}$ and Pf ≈ $10^{-11}$.

As can be seen, the best trade-off is obtained for a threshold setting S = 28. The simulation of the preamble detector shows isolated miss detection events. This observation can be used to further improve the reliability of the byte/frame synchronization. Indeed, the decision of a preamble loss can be taken when 3 successive PHY headers are not detected. A new byte/frame synchronization can be declared when the same correlators $C_k$ of each correlator-block indicate 3 successive preamble detections.

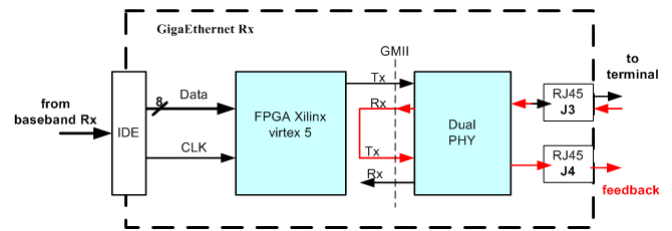

**Figure 13**. Gigabit Ethernet interface

Due to the low transmission rate for the return link, another system such as Wi-Fi or an Ethernet cable can be used.

## 4. Measurement results

A vector network analyzer (HP 8753D) was used to determine the frequency response and the impulse response of the both RF blocks (Tx + Rx) including LOS channel, using horn antennas. In this good propagation conditions, the measured response allows to estimate the system bandwidth and the RF components imperfections. The measurements were realized in a corridor with approximate dimensions of 15x2.4x3.1 m³ where the major part of the transmitted power is focused in the direction of the receiver. The RF-Tx and RF-Rx blocks were placed at a height of 1.5 m and at 10 m distance. After measurement set-up and calibration, a frequency response of 2 GHz bandwidth is obtained, as shown in Figure 14 (a).

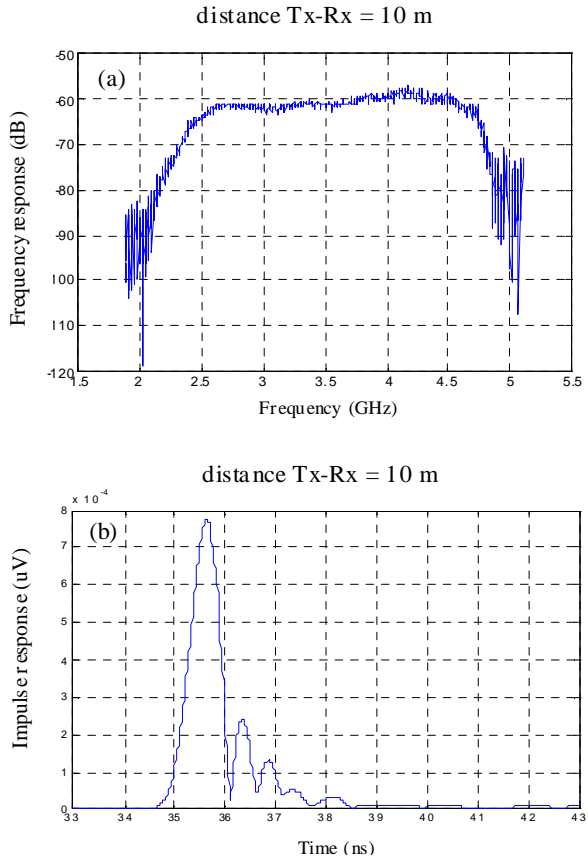

**Figure 14**. (a) Frequency response and (b) Impulse response

A perfect system must have an impulse response with only one lobe. As shown in Figure 14 (b), the measured impulse response presents some side lobes which are mainly due to RF components imperfections. A back-to-back test was also realized. The Tx and Rx antennas were used with a 45 dB fixed attenuator at 60 GHz. Similar results for the frequency and impulse responses were obtained, confirming the previous assumption.

Moreover, in order to evaluate the transmission performance, a HP70841B pattern generator was used to generate pseudo-random sequences at the transmitter and a HP 708842B error detector at the receiver. After demodulation, the eye pattern is shown in Figure 15. The eye opening indicates a good transmission quality at 875 Mbps for 10 m Tx-Rx distance.

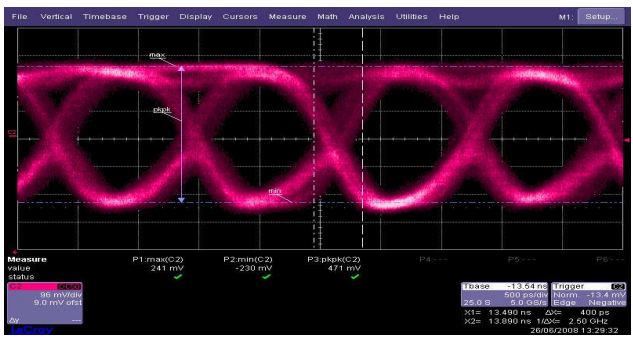

**Figure 15**. Eye diagram from 875 Mbps demodulated signal (10 m Tx-Rx distance)

From this eye pattern, the recovered data can be obtained by sampling at half-period in order to be noise resistant. Figure 16 shows the measured BER results at 875 Mbps as a function of the Tx-Rx distance. In our experiment, four antennas were used: two horn antennas and two patch antennas. Each patch antenna has 8 dBi gain with 30° HPBW.

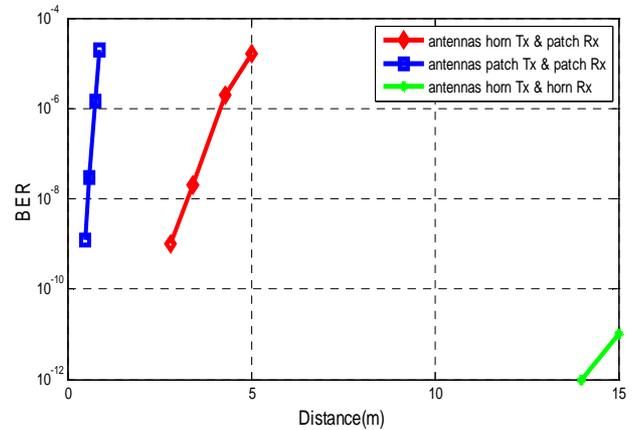

**Figure 16**. BER performance at 875 Mbps (without coding)

## 5. Discussion

The results show that the use of high-directivity horn antennas gives a remarkable BER performance and greatly reduces the harmful effects of the multipath propagation. This leads to a link budget improvement as high as 15 dB, compared to patch antennas. The system with high gain directive antennas is acceptable for point-to-point wireless communications links, with minimal multipath interference. However, a 60 GHz radio link is sensitive to shadowing due to high attenuation of the non line-of-sight (NLOS) propagation. Hence, when directive antennas are used, once the direct path is blocked by moving obstacles, the communication can be completely lost. Furthermore, the Tx-Rx antennas have to be aligned; otherwise the beam-pointing errors will cause an important reduction of the communication quality due to an increased BER. For LOS and point-to-point configurations, the use of a patch antenna for the receiver has the advantage to assure an important angular coverage, which significantly reduces the misalignment errors. However, if antennas beamwidth is large, equalization should be adopted to overcome the multi-path interference at a high data rate[10].

In order to improve the system reliability, the choice of a centralized transmitter, preferably located on the center of the room ceiling must be considered. The height of the transmitter (with less directional antenna) can partly reduce the influence of the shadowing. The different receivers forming the network could be equipped with more directional antennas, pointed toward the transmitter. It has been noted that the use of a directional receiving antenna is interesting for its gain and the reduction on the multipath propagation effects. In this configuration, only the contribution of the direct path is sought, and classical single carrier modulations could be used.

The previous measurements were performed in a corridor. In a large hall, the power of the received signal can be lower and thus significantly affects the quality of the communication link. Future work will provide BER measurement results (with channel coding) in different indoor environments.

## 6. Conclusion

This paper presented the design and the implementation of a 60 GHz communication system for WPAN applications in point-to-point or point-to-multipoint configurations. The proposed system provides a good trade-off between performance and complexity. An original method used for the byte and frame synchronization is also described. This method allows a high preamble detection probability and a very small false detection. For 1 Gbips reliable communications within large rooms, the transmitting and receiving antennas must have a relatively high gain.

Increasing the data rate (1,75 Gbps) can be achieved using higher order modulations such as QPSK. Implementation of frequency domain equalization methods is still under study. The system will be further enhanced to prove the feasibility of reliable wireless communications at data rates of several Gbps in different configurations, especially in non line-of-sight (NLOS) environments.

## Acknowledgements

This work is part of the research project Techim@ges supported by the French "Media & Network Cluster" and the COMIDOM project supported by the "Région Bretagne". The authors especially thank Guy Grunfelder (CNRS engineer) for his technical contributions during the system realization.